\documentclass[preprint,11pt,amsmath,amssymb,aps,showpacs,floatfix,prl,nofootinbib]{revtex4-1}

\usepackage{graphicx}
\usepackage{xcolor}
\usepackage{appendix}

\begin{document}

\title{Implications of primordial power spectra with statistical anisotropy on CMB temperature fluctuation and polarizations}

\author{Zhe Chang\footnote{E-mail: changz@ihep.ac.cn}}
\author{Sai Wang\footnote{E-mail: wangsai@ihep.ac.cn}}
\affiliation{Institute of High Energy Physics\\ Chinese Academy of Sciences, 100049 Beijing, China}


\begin{abstract}
Both the Wilkinson Microwave Anisotropy Probe (WMAP) and Planck observations reported the hemispherical asymmetry of the cosmic microwave background (CMB) temperature fluctuation. The hemispherical asymmetry might be stemmed from the primordial statistical anisotropy during the inflationary era of the universe. In this paper, we study possible implications of the primordial power spectra with dipolar anisotropy on the CMB temperature fluctuation and polarizations. We explicitly show that the statistical dipolar anisotropy may induce the off-diagonal (\(\ell'\neq\ell\)) $TT$, $EE$, $BB$, and $TE$ correlations, as well as the diagonal (\(\ell'=\ell\)) $TB$ and $EB$ spectra. In particular, these correlation coefficients are expected to be \(m\)-dependent generically. These signals of statistical anisotropy might be tested by CMB observations in future.
\end{abstract}
\maketitle

\section{I. Introduction}
\label{sec:Introduction}

Based on the Wilkinson Microwave Anisotropy Probe (WMAP) first-year data \cite{WMAP1},
it was found that the angular power spectrum of the cosmic microwave background (CMB) seems not to be isotropic at different positions on the celestial sphere \cite{hemispherical asymmetry01,hemispherical asymmetry02,hemispherical asymmetry03}.
The CMB power spectrum for one hemisphere was larger than that for the opposite hemisphere
at large angular scales with multipoles \(\ell\leq 40\).
Recently, the similar results were also shown by the WMAP nine-year data \cite{WMAP9} and the Planck 2013 result \cite{Planck2013resultsXXIII}.
It was found that the directions for the hemispherical asymmetry are close to each other for the WMAP-9 and Planck datasets.
The anisotropic amplitudes were also found to be around \(A\simeq 0.07\) for both observations.
Thus, the results of WMAP and Planck observations seem to be compatible with each other.
This might imply certain cosmic origin for the CMB hemispherical asymmetry.

The CMB hemispherical asymmetry implies that the universe is statistically anisotropic at large scales.
To be specific, the angular power spectrum of CMB temperature fluctuation could be characterized
by a dipolar modulation \cite{Spontaneous isotropy breaking,PrunetUzanetal2005,Gordon2007,WMAP717}.
It is well known that the CMB temperature fluctuation originates from the primordial perturbations in the very early universe.
Thus the anisotropic power spectrum might imply that the inflationary universe is intrinsically anisotropic.
Related to this, the anisotropic inflation could lead to dipole-modulated power spectra for the primordial scalar and tensor perturbations \cite{Inflation and primordial power spectra at anisotropic spacetime,Quadrupole--octopole alignment of CMB related to primordial power spectrum with dipolar modulation in anisotropic spacetime,Direction dependence of the power spectrum1302,Position Space CMB Anomalies from Multi-Stream Inflation,The CMB modulation from inflation,CMB dipole asymmetry from a fast roll phase,Small non-Gaussianity and dipole asymmetry in the CMB,Large Scale Anisotropic Bias from Primordial non-Gaussianity,Hemispherical Asymmetry and Local non-Gaussianity a Consistency Condition,Asymmetric Sky from the Long Mode Modulations,A Hemispherical Power Asymmetry from Inflation,Obtaining the CMB anomalies with a bounce from the contracting phase to inflation,CMB anomalies from an inflationary model in string theory,CMB Power Asymmetry from Primordial Sound Speed Parameter,Isocurvature and Curvaton Perturbations with Red Power Spectrum and Large Hemispherical Asymmetry,Hemispherical Power Asymmetry from Scale-Dependent Modulated Reheating,Cosmic microwave background anomalies in an open universe,Large-scale anomalies from primordial dissipation,Scale-dependent CMB asymmetry from primordial configuration,A viable explanation of the CMB dipolar statistical anisotropy,CMB Power Asymmetry from Non-Gaussian Modulation,Koivisto:2008xf}.
Besides the hemispherical asymmetry, these primordial statistical anisotropy
might leave more imprints on the CMB temperature fluctuation and polarizations.

In the traditional inflation model, the assumption of statistical isotropy
guarantees the prediction that different multipole moments are uncorrelated
and the CMB power spectra are \(m\)-independent.
Thus the power spectra have the diagonal form as \(\langle a_{X,\ell m}a^{\ast}_{X',\ell' m'}\rangle=C_{XX',\ell}\delta_{\ell\ell'}\delta_{mm'}\), where \(X\) denotes \(T\), \(E\) and \(B\), respectively.
In addition, the $TB$ and $EB$ spectra vanish since the $B$-mode polarization has opposite parity
to the temperature fluctuation and $E$-mode polarization \cite{All-sky analysis of polarization in the microwave background,Zaldarriaga1997a,Kamionkowski1997a,polarization tensor field by Kamionkowski}.
In the case of statistical anisotropy, however, the off-diagonal correlations (including $TB$ and $EB$ correlations) would arise
and depend on \(m\)-components \cite{Imprints of the anisotropic inflation on the cosmic microwave background,Ma1102,CMB statistics for a direction-dependent primordial power spectrum,Cosmological Signature of New Parity-Violating Interactions,Testable polarization predictions for models of CMB isotropy anomalies,CMB Anomalies from Relic Anisotropy,Inflationary perturbations in anisotropic backgrounds2007,Bipolar Harmonic encoding of CMB correlation patterns,Large-angle anomalies in the CMB,Imprints of a Primordial Preferred Direction on the Microwave Background,Effect on cosmic microwave background polarization of coupling of quintessence to pseudoscalar formed from the electromagnetic field and its dual,Measuring Statistical isotropy of the CMB anisotropy,Boehmer:2007ut}.
This reveals that there is mixing of power between different multipole moments.
The diagonal $TB$ and $EB$ spectra could also appear in condition of the parity breaking.
Therefore, the statistical anisotropy together with its origination could be tested by the off-diagonal correlations (\(\ell'\neq \ell\))
and the cross-correlations between $T$ and $B$, or $E$ and $B$.
Actually, Liu \& Li \cite{Abnormal correlation in WMAP data} has found clues that there are abnormal correlations in the WMAP data.
However, their conclusion is not conclusive since they could not determine whether the abnormal correlations originate from the parity violation or the data flaw.

In this paper, we study possible implications of the primordial statistical anisotropy on the CMB temperature fluctuation and polarizations. For the primordial power spectra with dipole modulation, the CMB angular correlation coefficients are calculated explicitly,
including the diagonal and off-diagonal correlations.
Their \(m\)-dependence will be shown as well.
The mixing of power between different multipoles would be quantified.
The $TB$ and $EB$ spectra would be nonzero.
We would point out certain predicted signals for the anisotropic inflation.
The rest of the paper is arranged as follows.
In section II\ref{sec:CMB correlation}, we calculate the angular correlation coefficients of the CMB temperature fluctuation and polarizations in the generic case of statistical anisotropy.
For the primordial power spectra with a dipole modulation, their imprints on the CMB physics would be explicitly revealed in section III \ref{sec:Dipolar anisotropy}.
Conclusions and discussions are listed in section IV\ref{sec:Conclusion}.

\section{II. CMB correlation coefficients: Statistical anisotropy}
\label{sec:CMB correlation}

Generically, the primordial statistical anisotropy could be accounted by the anisotropic power spectra of initial scalar and tensor perturbations.
There is a convenient way to express the primordial power spectra with anisotropy as
\begin{equation}
\label{anisotropic primordial power spectra}
\langle\zeta^{s}(\textbf{k})\zeta^{s'\ast}(\textbf{k}')\rangle=\frac{2\pi^2}{k^3}P^{ss'}(\textbf{k})\delta^{(3)}(\textbf{k}-\textbf{k}')\ ,
\end{equation}
where \(s=0\) denotes the scalar perturbation and \(s=\pm2\) denotes the tensor perturbation with polarization \(\pm2\), respectively.
In the isotropic inflation model, the above spectra depend only on the length of \(\textbf{k}\)-wavevector,
and thus they are blind for spatial directions.
However, the statistical anisotropy induces the direction dependence to the primordial spectra.

The CMB temperature fluctuation and polarizations are usually expanded in terms of the spherical harmonics, i.e.,
\begin{equation}
X(\hat{\textbf{p}},\textbf{x}_0)=\sum_{\ell=2}^{\infty}\sum_{m=-\ell}^{\ell}a_{X,\ell m}Y_{\ell m}(\hat{\textbf{p}};\hat{\textbf{e}})\ ,
\end{equation}
where \(X\) denotes \(T\), \(E\) and \(B\), respectively.
The spherical function \(Y_{\ell m}(\hat{\textbf{p}};\hat{\textbf{e}})\) is measured from the unit vector \(\hat{\textbf{e}}\).
By inversing the above expression, we could obtain the expansion coefficients \(a_{X,\ell m}\) as
\begin{equation}
a_{X,\ell m}=\int d\Omega_{\hat{\textbf{p}}}X(\hat{\textbf{p}},\textbf{x}_0)Y^\ast_{\ell m}(\hat{\textbf{p}};\hat{\textbf{e}})\ .
\end{equation}
In this way, the angular correlation coefficients are given by
\begin{equation}
\label{angular correlation coefficients}
\langle a_{X,\ell m}a^{\ast}_{X',\ell' m'}\rangle=C_{XX',\ell\ell', mm'}\ .
\end{equation}
They are not diagonal in the case of statistical anisotropy, since \(a_{X,\ell m}\) depend on \(\hat{\textbf{e}}\).

On the other hand, the CMB temperature fluctuation and polarizations are evolved from the primordial scalar and tensor perturbations
\(\zeta^{s=0,\pm2}\) generated in the inflation phase of the universe.
Thus the fluctuations \(X\) could be expressed as
\begin{equation}
\label{fluctuation X from primordial fluctuation}
X(\hat{\textbf{p}},\textbf{x}_0)=\int \frac{d^3\textbf{k}}{(2\pi)^3}\sum_{\ell}\sum_{s=-2}^{2}\zeta^{s}(\textbf{k},\eta_i)Y_{\ell s}(\hat{\textbf{p}};\hat{\textbf{k}})
\Delta_{X,\ell s}(k)e^{i\textbf{k}\cdot \textbf{x}_0}\ ,
\end{equation}
where \(\Delta_{X,\ell s}(k)\) denote the transfer functions.
Here the orientation of spherical harmonics \(Y_{\ell s}(\hat{\textbf{p}};\hat{\textbf{k}})\)
is determined by the unit vector \(\hat{\textbf{k}}\), which is along the propagating direction of initial perturbations.

From Eqs.~(\ref{anisotropic primordial power spectra})--(\ref{fluctuation X from primordial fluctuation}),
the CMB angular correlation coefficients could be obtained as follows
\begin{equation}
\label{CXXllmm}
C_{XX',\ell\ell', mm'}=\int\frac{d\ln k}{(2\pi)^3}\sum_{s,s'}\Delta_{X,\ell s}(k)\Delta^\ast_{X',\ell's'}(k)P^{ss'}_{\ell\ell'mm'}\ ,
\end{equation}
where
\begin{equation}
\label{Pssllmm}
P^{ss'}_{\ell\ell'mm'}\equiv\int d\Omega_{\hat{\textbf{k}}}P^{ss'}(\textbf{k})_{-s}Y^\ast_{\ell m}(\hat{\textbf{k}};\hat{\textbf{e}})_{-s'}Y_{\ell' m'}(\hat{\textbf{k}};\hat{\textbf{e}})\ ,
\end{equation}
and an extra factor \((2\ell+1)^{-1/2}\) has been absorbed into the transfer function \(\Delta_{X,\ell s}(k)\).
Here \(_{s}Y_{\ell m}\) are the spin-\(s\) weighted spherical harmonics and \(_{0}Y_{\ell m}\) is just \(Y_{\ell m}\) \cite{All-sky analysis of polarization in the microwave background}.
They are used to expand the Stokes parameters of the CMB temperature fluctuation and polarizations.
In the above derivation, we have used a relation of the form
\begin{equation}
Y_{\ell s}(\hat{\textbf{p}};\hat{\textbf{k}})=\sqrt{\frac{4\pi}{2\ell+1}}\sum_{m}Y_{\ell m}(\hat{\textbf{p}};\hat{\textbf{e}})~_{-s}Y^{\ast}_{\ell m}(\hat{\textbf{k}};\hat{\textbf{e}})\ .
\end{equation}
One could refer to Ref.~\cite{Imprints of the anisotropic inflation on the cosmic microwave background} for a similar derivation of the CMB correlations.

In the isotropic case, the primordial power spectra (\ref{anisotropic primordial power spectra}) would be irrelative to the spatial directions.
Thus \(P^{ss'}_{\ell\ell'mm'}\) would become diagonal (\(\propto\delta_{\ell\ell'}\delta_{m m'}\))
according to the orthogonality relation of \(_{s}Y_{\ell m}\).
The correlation coefficients (\ref{CXXllmm}) reduce back to the conventional power spectra.
In the anisotropic case, by contrast, the correlation coefficients (\ref{CXXllmm}) would be off-diagonal
because of the direction dependence of the primordial spectra (\ref{anisotropic primordial power spectra}).
In addition, the $TB$ and $EB$ correlations would also arise.
In particular, they could be diagonal only in a special case.
In the following section, we would consider an explicit form of primordial power spectra with dipolar anisotropy
which might be related to the CMB hemispherical asymmetry.

\section{III. CMB correlation coefficients: Dipolar anisotropy case}
\label{sec:Dipolar anisotropy}

In the recent work by Chang \& Wang \cite{Inflation and primordial power spectra at anisotropic spacetime,Quadrupole--octopole alignment of CMB related to primordial power spectrum with dipolar modulation in anisotropic spacetime},
we proposed an anisotropic inflation model to account for the CMB hemispherical asymmetry
as well as to relieve the issue of alignment between the CMB low-\(\ell\) multipole moments.
This inflation model is based on the Randers-Finsler spacetime \cite{Randers space},
where the anisotropy is induced by a 1-form in the Randers structure.
Finsler geometry \cite{Book by Rund,Book by Bao} is a natural framework to deal with the issue of spacetime anisotropy.
There are less symmetries in the Finsler spacetime than those in the Riemann spacetime \cite{Finsler isometry by Wang,Finsler isometry by Rutz,Finsler isometry LiCM,On relativistic symmetry of Finsler spaces with mutually opposite preferred directions}.
Thus the Finsler spacetime is intrinsically anisotropic.
In the Randers-inflation model, we explicitly showed that the primordial power spectrum of scalar perturbation is modulated
by a spatial dipole.
In addition, it is compatible with the constraint on the anisotropic amplitude from the distribution of quasars
\cite{Constraints on cosmic hemispherical power anomalies from quasars}.

In the Randers spacetime, the forward geodesic is different from the backward one.
The parity $P$ is broken.
The primordial perturbation with wavevector \(\textbf{k}\) would propagate differently from the one with wavevector \(-\textbf{k}\).
Thus the primordial power spectra are given by \cite{Inflation and primordial power spectra at anisotropic spacetime,Quadrupole--octopole alignment of CMB related to primordial power spectrum with dipolar modulation in anisotropic spacetime}
\begin{equation}
\label{primordial power spectra}
P^{00,\pm2\pm2}(\textbf{k})=P_{iso}^{00,\pm2\pm2}(k)\left(1+\frac{k_c}{k}\hat{\textbf{k}}\cdot \hat{\textbf{n}}\right)\ ,
\end{equation}
where the upper labels $00$ and $\pm2\pm2$ denote the initial scalar and tensor perturbations, respectively.
The unit vector \(\hat{\textbf{n}}\) denotes a privileged direction in the cosmic space.
The constant \(k_{c}\) is a typical wavenumber which refers to the scale of statistical anisotropy.
Here the power spectrum for primordial tensor perturbation is assumed,
while it could be derived via the osculating Riemannian method \cite{Book by Rund,Book by Asanov,FRW model with weak anisotropy by Stavrinos} in the Randers spacetime.
One should note that the primordial power spectra with dipolar anisotropy are also proposed in other inflation models \cite{Direction dependence of the power spectrum1302,Position Space CMB Anomalies from Multi-Stream Inflation,The CMB modulation from inflation,CMB dipole asymmetry from a fast roll phase,Small non-Gaussianity and dipole asymmetry in the CMB,Large Scale Anisotropic Bias from Primordial non-Gaussianity,Hemispherical Asymmetry and Local non-Gaussianity a Consistency Condition,Asymmetric Sky from the Long Mode Modulations,A Hemispherical Power Asymmetry from Inflation,Obtaining the CMB anomalies with a bounce from the contracting phase to inflation,CMB anomalies from an inflationary model in string theory,CMB Power Asymmetry from Primordial Sound Speed Parameter,Isocurvature and Curvaton Perturbations with Red Power Spectrum and Large Hemispherical Asymmetry,Hemispherical Power Asymmetry from Scale-Dependent Modulated Reheating,Cosmic microwave background anomalies in an open universe,Large-scale anomalies from primordial dissipation,Scale-dependent CMB asymmetry from primordial configuration,A viable explanation of the CMB dipolar statistical anisotropy,CMB Power Asymmetry from Non-Gaussian Modulation}.

For the primordial tensor perturbation, we could obtain the CMB correlation coefficients as follows
\begin{equation}
\label{tensor}
C_{XX',\ell\ell', mm'}^{T}=\int\frac{d\ln k}{(2\pi)^3}\Delta_{X,\ell 2}(k)\Delta^\ast_{X',\ell'2}(k)P^{\pm2\pm2}_{\ell\ell'mm'}\ ,
\end{equation}
where
\begin{eqnarray}
\label{p+-2}
P^{\pm2\pm2}_{\ell\ell'mm'}&&=\int d\Omega~ P^{\pm2\pm2}\left(_{-2}Y^\ast_{\ell m}~_{-2}Y_{\ell' m'}\pm~_{+2}Y^\ast_{\ell m}~_{+2}Y_{\ell' m'}\right)\nonumber\\
&&=P^{\pm2\pm2}_{iso}\delta_{mm'}\left(2\delta_{\ell\ell'}+\sqrt{\frac{2\ell'+1}{2\ell+1}}\frac{k_c}{k}\mathcal{C}^{\ell' m}_{10\ell m}\left(\mathcal{C}^{\ell' 2}_{10\ell 2}\pm\mathcal{C}^{\ell' -2}_{10\ell -2}\right)\right)\ ,
\end{eqnarray}
and \(\mathcal{C}^{\ell m}_{L M \ell' m'}\) is the Clebsch-Gordan coefficient.
Here we choose ``$+$'' for $TT$, $EE$, $BB$, $TE$ correlations and ``$-$'' for $TB$, $EB$ correlations in the above expression.
To get the above results, we have used the relations \(\Delta_{\ell, -2}=\Delta_{\ell, 2}\) for $T$ and $E$,
and \(\Delta_{\ell, -2}=-\Delta_{\ell, 2}\) for $B$.
In addition, there is an essential relation \cite{Imprints of the anisotropic inflation on the cosmic microwave background}
\begin{equation}
\int d\Omega~ Y_{L M}~_{-s}Y^\ast_{\ell m}~_{-s}Y_{\ell' m'}=\sqrt{\frac{(2L+1)(2\ell'+1)}{4\pi (2\ell+1)}}\mathcal{C}^{\ell m}_{L M \ell' m'}\mathcal{C}^{\ell s}_{L 0 \ell' s}\ ,
\end{equation}
which was used at the last step of Eq.~(\ref{p+-2}).

In Eq.~(\ref{p+-2}), the diagonal part (\(\propto \delta_{\ell \ell'}\)) corresponds to the statistically isotropic contribution.
The remaining off-diagonal part results from the contribution of statistical anisotropy.
From Eqs.~(\ref{tensor}) and (\ref{p+-2}), we know that the correlation coefficients are proportional to \(\delta_{m m'}\).
The reason is that the primordial power spectrum is axisymmetric along the privileged axis \(\hat{\textbf{n}}\)
for the tensor perturbation (\ref{primordial power spectra}).
Even so, the correlation coefficients are \(m\)-dependent.
The reason is that the parity \(P\) is broken along this privileged direction.
This could be revealed by the factor \(\mathcal{C}^{\ell' m}_{10\ell m}\) in Eq.~(\ref{p+-2}).
For different \(m\), it would not always equal to the same value.
Similar results could be obtained for the scalar perturbation as will be discussed below.

For the $TT$, $EE$, $BB$ and $TE$ correlations, Eq.~(\ref{tensor}) would become diagonal unless the condition \(\ell'=\ell\pm1\) is satisfied.
The off-diagonal correlations arise here.
Thus there is mixing of power between two neighbor multipoles.
For the $TB$ and $EB$ correlations, by contrast, Eq.~(\ref{tensor}) would vanish unless \(\ell'=\ell\).
Thus the diagonal $TB$ and $EB$ spectra appear.
To calculate Eq.~(\ref{p+-2}), we have used the symmetry of Clebsch-Gordan coefficient
\(\mathcal{C}^{\ell' m'}_{L M \ell m}=(-1)^{L+\ell-\ell'}\mathcal{C}^{\ell' -m'}_{L -M \ell -m}\).
The above results reveal the cosmic manifestations of parity breaking for the CMB physics.
In addition, the statistically anisotropic contribution is significant for small \(k\) only which refers to large scales.

For the primordial scalar perturbation, we calculate the $TT$, $EE$, and $TE$ correlation coefficients as follows
\begin{equation}
\label{scalar}
C_{XX',\ell\ell', mm'}^{S}=\int\frac{d\ln k}{(2\pi)^3}\Delta_{X,\ell 0}(k)\Delta^\ast_{X',\ell'0}(k)P^{00}_{\ell\ell'mm'}\ ,
\end{equation}
where
\begin{eqnarray}
\label{p00}
P^{00}_{\ell\ell'mm'}&&=\int d\Omega~ P^{00}~Y^\ast_{\ell m}~Y_{\ell' m'}\nonumber\\
&&=P^{00}_{iso}\delta_{mm'}\left(\delta_{\ell\ell'}+\sqrt{\frac{2\ell'+1}{2\ell+1}}\frac{k_c}{k}\mathcal{C}^{\ell' m}_{1 0 \ell m}\mathcal{C}^{\ell' 0}_{1 0 \ell 0}\right)\ .
\end{eqnarray}
Other correlation coefficients would vanish since the scalar perturbation does not generate the B-mode polarization, namely, \(\Delta_{\ell, 0}=0\).

\begin{figure}[h]
\begin{center}
\includegraphics[width=8 cm]{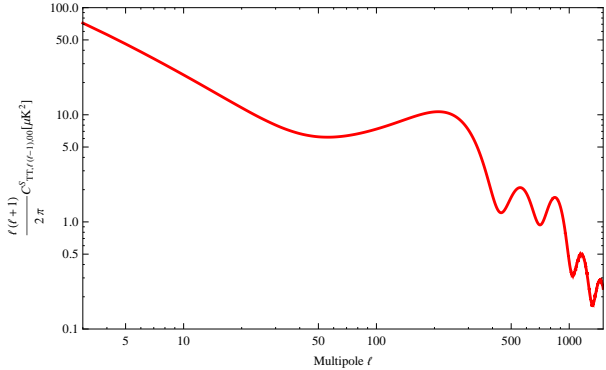}
\caption{[Color online] The off-diagonal TT correlation coefficients \(\frac{\ell(\ell+1)}{2\pi}C^{S}_{TT,\ell(\ell-1),00}\) contributed by statistical anisotropy. The typical parameter \(k_c\) is chosen as \(k_{c}^{-1}\simeq 14~\rm{Gpc}\) and the WMAP-1 data \cite{WMAP1} is used.}
\label{ScalarTTll-100}
\end{center}
\end{figure}

\begin{figure}[h]
\begin{center}
\includegraphics[width=8 cm]{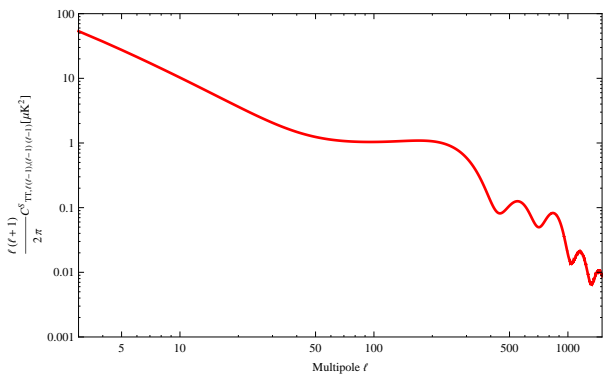}
\caption{[Color online] The off-diagonal TT correlation coefficients \(\frac{\ell(\ell+1)}{2\pi}C^{S}_{TT,\ell(\ell-1),(\ell-1)(\ell-1)}\) demonstrate the \(m\)-dependence of the correlations contributed by the statistical anisotropy. Here the cosmic parameters are chosen as those in Fig.~\ref{ScalarTTll-100}.}
\label{ScalarTTll-1l-1l-1}
\end{center}
\end{figure}

Similar to discussions on the tensor perturbation, we could obtain the implications of
the scalar perturbation with statistical anisotropy on the CMB physics as follows.
The diagonal part (\(\propto \delta_{\ell \ell'}\)) in Eq.~(\ref{p00})
refers to the contribution of statistical isotropy, while the off-diagonal part refers to that of statistical anisotropy.
The Eq.~(\ref{scalar}) would become diagonal unless the condition \(\ell'=\ell\pm1\) is satisfied.
Thus there is mixing of power between different multipoles as expected.
To be specific, we depict the contribution of statistical anisotropy to the off-diagonal TT correlation coefficients
\(\frac{\ell(\ell+1)}{2\pi}C^{S}_{TT,\ell(\ell-1),00}\) in Fig.~\ref{ScalarTTll-100}.
The WMAP-1 data \cite{WMAP1} is used and the typical parameter \(k_c\) is chosen as \(k_{c}^{-1}\simeq 14~\rm{Gpc}\)
which is related to the last-scattering surface.
Here the off-diagonal correlations are remained between two neighbor multipoles.
However, they are decayed with the increase of \(\ell\).
The reason is that the statistical anisotropy is just significant at large scales for the primordial spectra (\ref{primordial power spectra}).
One should note that this decay is model dependent.
In addition, the correlation coefficients are proportional to \(\delta_{m m'}\) for the axisymmetry along the privileged axis
\(\hat{\textbf{n}}\) for the scalar spectra (\ref{primordial power spectra}).
However, they are still \(m\)-dependent since there is \(\mathcal{C}^{\ell' m}_{10\ell m}\) in Eq.~(\ref{p00}).
To be specific, we depict the off-diagonal TT correlation \(\frac{\ell(\ell+1)}{2\pi}C^{S}_{TT,\ell(\ell-1),(\ell-1)(\ell-1)}\)
in Fig.~\ref{ScalarTTll-1l-1l-1}.
Here we still choose \(k_{c}^{-1}\simeq 14~\rm{Gpc}\) and WMAP-1 data.
By comparing Fig.~\ref{ScalarTTll-1l-1l-1} with Fig.~\ref{ScalarTTll-100},
we could find that the correlation between two neighbor multipoles with \(m=0\) is different from that with \(m=\ell-1\).
This is a generic result for the CMB angular correlations arising from the statistical anisotropy.

\section{IV. Conclusions and discussions}
\label{sec:Conclusion}

In this paper, we have studied possible implications of the primordial statistical anisotropy on the CMB temperature fluctuation and polarizations. In general, the statistical anisotropy could give rise to the off-diagonal correlations between different multipole moments. These correlations would be \(m\)-dependent. In particular, we explicitly analyzed the cosmological implications of primordial power spectra with dipolar anisotropy. It was found that there could be diagonal $TB$ and $EB$ spectra as well as the on/off-diagonal $TT$, $EE$, $BB$ and $TE$ correlations. The diagonal $TT$, $EE$, $BB$ and $TE$ spectra arise from the statistically isotropic contribution. By contrast, their off-diagonal counterparts and the diagonal $TB$, $EB$ spectra originate from the statistical anisotropy with a spatial dipole. We depicted in Fig.~\ref{ScalarTTll-100} the off-diagonal $TT$ correlation coefficients \(\frac{\ell(\ell+1)}{2\pi}C^{S}_{TT,\ell(\ell-1),00}\) to quantify the level of statistical anisotropy,
which is of level \(\sim 0.1\) at large scales.
Considering the cosmic variance, one should be able to test the statistical anisotropy at the scales $\ell \sim 100$.
In addition, we plotted in Fig.~\ref{ScalarTTll-1l-1l-1} the off-diagonal $TT$ correlation coefficients \(\frac{\ell(\ell+1)}{2\pi}C^{S}_{TT,\ell(\ell-1),(\ell-1)(\ell-1)}\) to demonstrate the \(m\)-dependence of correlations contributed by the statistical anisotropy. Our analysis could be generalized to more complex forms of statistical anisotropy.

We could be able to test the predicted cosmological implications of statistical anisotropy via the recent/future CMB observations in principle. Note that it is difficult to test the statistical anisotropy which depends on the spatial scale such as the primordial spectra (\ref{primordial power spectra}). However, a recent work presented a way to test the dipolar modulation without the scale dependence \cite{Testing the Dipole Modulation Model in CMBR}. This corresponds to the case that we fix \(k\) in (\ref{primordial power spectra}) as a constant scale. They used the WMAP and Planck data to study the off-diagonal TT correlations while they found no significant signals for the dipolar modulation. There are several reasons for their result. One reason refers to that the dipolar anisotropy has the scale dependence. As was mentioned above, however, it is difficult to deal with this scale dependence. The other reason is that the anisotropy is not just a dipolar form. This may relate to even more complex anisotropic models as they have discussed \cite{Testing the Dipole Modulation Model in CMBR}. Another reason is that the statistical anisotropy is beyond the sensitivity of the WMAP and Planck satellites. Maybe future CMB observations could give more stringent constraints on the statistical anisotropy of the universe.

One may propose that the primordial power spectra (\ref{primordial power spectra}) cannot be realized \cite{Referee's comments}.
Actually, this proposition is true just in the framework of Riemann geometry.
To be specific, the Fourier transformation for a real field \(\phi(\textbf{x})\) is given by
\begin{equation}
\label{Fourier transformation 11}
\phi(\textbf{k})=\int d^3\textbf{x} \phi(\textbf{x})e^{-i\textbf{k}\cdot\textbf{x}}\ ,
\end{equation}
which implies a relation
\begin{equation}
\label{relation 11}
\phi(-\textbf{k})=\phi^\ast(\textbf{k})\ .
\end{equation}
Here, \(\textbf{k}\cdot\textbf{x}\) is the Euclidean inner product between three dimensional vectors \(\textbf{k}\) and \(\textbf{x}\).
Noting that the power spectrum is the amplitude square of the field \(\phi(\textbf{k})\), one obtains
\begin{equation}
\label{power spectrum 11}
P_{\phi}(\textbf{k})=P_{\phi}(-\textbf{k})\ .
\end{equation}
The above equation implies that the parameter \(k_c\) in (\ref{primordial power spectra}) must vanish.
In this way, the primordial power spectra in (\ref{primordial power spectra}) are reduced back to the isotropic case.

In Finsler geometry, nevertheless, the above demonstration is invalid generally.
The reason is that Finsler space gets rid of the quadratic restriction on the spatial interval.
Finsler geometry \cite{Book by Rund,Book by Bao} is a natural framework to deal with the issue of anisotropy.
There are less symmetries in the Finsler space than those in the Riemann space \cite{Finsler isometry by Wang,Finsler isometry by Rutz,Finsler isometry LiCM,On relativistic symmetry of Finsler spaces with mutually opposite preferred directions}.
Thus, the Finsler space is intrinsically anisotropic.
The Finsler spatial interval is given by
\begin{equation}
ds^2=g_{ij}(x,dx)dx^idx^j\ .
\end{equation}
Related to this, the plane wave in a flat Finsler space could be given by
\begin{equation}
\label{plane wave in Finsler 11}
\psi_{\textbf{k}}(\textbf{x})\propto e^{ig_{mn}(\textbf{k})k^mx^n}\ ,
\end{equation}
where \(k^i:=dx^i/ds\) is the wavevector.
The Finslerian Fourier transformation for the real field \(\phi(\textbf{x})\) is given by
\begin{equation}
\label{Finslerian Fourier transformation 11}
\phi(\textbf{k})=\int d^3\textbf{x} \phi(\textbf{x})e^{ig_{mn}(\textbf{k})k^mx^n}\ .
\end{equation}
In this way, the relation (\ref{relation 11}) may be no longer held, namely,
\begin{equation}
\label{Finsler relation 11}
\phi(-\textbf{k})\neq\phi^\ast(\textbf{k})\ .
\end{equation}
The reason is that the Finsler metric \(g_{mn}(\textbf{x},\textbf{k})\) could change its sign
under the transformation \(\textbf{k}\rightarrow-\textbf{k}\), namely,
\begin{equation}
\label{gneq-g}
g_{mn}(\textbf{x},\textbf{k})\neq g_{mn}(\textbf{x},-\textbf{k})\ .
\end{equation}
For instance, the Randers metric is given by \cite{Randers space}
\begin{equation}
ds=\sqrt{\tilde{a}_{mn}(\textbf{x})dx^mdx^n}+\tilde{b}_m(\textbf{x}) dx^m\ ,
\end{equation}
where \(\tilde{a}_{mn}\) is a Riemann metric and \(\tilde{b}_m dx^m\) is a 1--form.
Under the transformation \(dx^i\rightarrow-dx^i\), the Riemannian part is invariant while the 1--form changes its sign.
One could check that this property will result in the consequence (\ref{gneq-g}).
Therefore, the primordial power spectra (\ref{primordial power spectra}) can be realized in Finsler geometry.

\vspace{0.5 cm}
\begin{acknowledgments}
We are grateful to Prof. Xin Li, Dr. Si-Yu Li, Hai-Nan Lin and Dong Zhao for useful discussions.
One of us (S.W.) thanks Prof. Rong-Gen Cai and Zong-Kuan Guo for their hospitality
at the Institute of Theoretical Physics, Chinese Academy of Sciences.
This work has been funded by the National Natural Science Fund of China under Grant No. 11075166 and No. 11147176.
\end{acknowledgments}

\end{document}